\documentclass[twocolumn,showpacs,amsmath,amssymb]{revtex4}
\usepackage{graphicx}
\begin{document}
\title{Hamiltonian formalism in Friedmann cosmology and its quantization}
\author{Jie Ren$^1$}
\email{jrenphysics@hotmail.com}
\author{Xin-He Meng$^{2,3}$}
\author{Liu Zhao$^2$}
\affiliation{$^1$Theoretical Physics Division, Chern Institute of
Mathematics, Nankai University, Tianjin 300071, China}
\affiliation{$^2$Department of physics, Nankai University, Tianjin
300071, China} \affiliation{$^3$BK21 Division of Advanced Research
and Education in physics, Hanyang University, Seoul 133-791, Korea}
\date{\today}
\begin{abstract}
We propose a Hamiltonian formalism for a generalized
Friedmann-Roberson-Walker cosmology model in the presence of both a
variable equation of state (EOS) parameter $w(a)$ and a variable
cosmological constant $\Lambda(a)$, where $a$ is the scale factor.
This Hamiltonian system containing 1 degree of freedom and without
constraint, gives Friedmann equations as the equation of motion,
which describes a mechanical system with a variable mass object
moving in a potential field. After an appropriate transformation of
the scale factor, this system can be further simplified to an object
with constant mass moving in an effective potential field. In this
framework, the $\Lambda$ cold dark matter model as the current
standard model of cosmology corresponds to a harmonic oscillator. We
further generalize this formalism to take into account the bulk
viscosity and other cases. The Hamiltonian can be quantized
straightforwardly, but this is different from the approach of the
Wheeler-DeWitt equation in quantum cosmology.
\end{abstract}
 \maketitle

\section{Introduction}
Since the current accelerating expansion of our Universe was
discovered \cite{de} around 1998 and 1999, theoretical physicists
have devoted increasingly more attention to the
Friedmann-Roberson-Walker (FRW) model as a standard framework in
cosmology study. The $\Lambda$ cold dark matter ($\Lambda$CDM) model
as the standard model of cosmology so far fits well with
observational data whereas it has had some serious theoretical
problems. To make a comparison to the $\Lambda$CDM model, physicists
have built many cosmological models that are able to give out the
effective Friedmann equations with variable cosmological constant.
To quantize the Friedmann equations, the commonly used theory is the
Wheeler-DeWitt equation \cite{dewitt}, which has been studied and
applied widely in quantum cosmology \cite{wdw}. Starting from the
Hilbert-Einstein action with the Roberson-Walker (RW) metric, the
corresponding Hamiltonian $\mathcal{H}$ can be obtained. Then the
Friedmann equation plays the role as the constraint $\mathcal{H}=0$,
which leads to the Wheeler-DeWitt equation. In the present work, we
consider the Friedmann equations as basic equations and find a
Hamiltonian system that gives Friedmann equations as classical
equations of motion without constraint.

Many Ans\"{a}tze of the variable cosmological constant have been
studied in the literature, such as Refs.~\cite{vcc,vcc1,vcc2,wm}.
Moreover, some models motivated from the string theory give an
effective cosmological term when reduced to the FRW framework. We
assume that the equation of state (EOS) parameter $w\equiv p/\rho$
can also be variable, which means that the contents of the Universe,
except the cosmological term, are generalized to a nonperfect fluid,
or perfect fluid as a special case. In observational cosmology, the
redshift $z$ is regarded as an observable quantity and related to
the scale factor $a$ by $z\equiv a_0/a-1$. Therefore, we investigate
a general case that both the EOS parameter $w$ and the cosmological
constant $\Lambda$ can be functions of the scale factor $a$, and
take into account the bulk viscosity.

As an extension of the problem, we construct a Hamiltonian formalism
for a system described by the following equation:
$$\ddot{q}=f_1(q)\dot{q}^2+\eta\dot{q}+f_2(q),$$
where $f_1(q)$ and $f_2(q)$ are arbitrary functions, and $\eta$ is
constant. Also it can be regarded as a generalization of the damping
harmonic oscillator. The corresponding Hamiltonian describes an
object with variable mass moving in a potential field. After an
appropriate canonical transformation, this system can be further
simplified to an object with constant mass moving in an effective
potential field. Thus, differential models in the FRW framework are
characterized by their effective potentials. This is a general
formalism and it can be applied to many cosmological models, for
example, that the $\Lambda$CDM model corresponds to a harmonic
oscillator. Since the quantization of Friedmann equations can
provide an insight to quantum cosmology as a glimpse of quantum
gravity, we also make some remarks on the quantum case, which
provides a correspondence between cosmology and quantum mechanics.

The paper is organized as follows. In Sec. II we present a
generalized FRW model and the corresponding Hamiltonian to describe
the Friedmann equations. Then we find a canonical transformation to
further simplify the problem, and give some examples and special
cases. In Sec. III we show that our framework can also be applied in
the dissipative case with bulk viscosity. In Sec. IV we turn our
attention to the relation to the observable quantities and review
some issues of the Bianchi identity. In Sec. V we make some remarks
on quantum cosmology from our approach. In the last section we
present the conclusion and discuss some future subjects.

\section{Hamiltonian formalism}
\subsection{Hamiltonian description of the Friedmann equations}
We consider the RW metric in the flat space geometry ($k$=0) as the
case favored by current cosmic observational data:
\begin{equation}
ds^2=-dt^2+a(t)^2(dr^2+r^2d\Omega^2),
\end{equation}
where $a(t)$ is the scale factor. The energy-momentum tensor for the
cosmic fluid can be written as
\begin{equation}
\tilde{T}_{\mu\nu}=(\rho+p)U_\mu
U_\nu+(p+\rho_\Lambda)g_{\mu\nu},\label{eq:tmn}
\end{equation}
where $\rho_\Lambda=\Lambda/(8\pi G)$ is the energy density of the
cosmological constant. Thus, Einstein's equation
$R_{\mu\nu}-\frac{1}{2}g_{\mu\nu}R=8\pi G\tilde{T}_{\mu\nu}$
contains two independent equations:
\begin{subequations}
\begin{eqnarray}
\frac{\dot{a}^2}{a^2} &=& \frac{8\pi G}{3}\rho+\frac{\Lambda}{3}\label{eq1},\\
\frac{\ddot{a}}{a} &=& -\frac{4\pi
G}{3}(\rho+3p)+\frac{\Lambda}{3}\label{eq2}.
\end{eqnarray}
\end{subequations}
The EOS of the matter (cosmic fluid except the cosmological
constant) is commonly assumed to be
\begin{equation}
p=(\gamma-1)\rho.
\end{equation}
Cosmologists usually call Eq.~(\ref{eq1}) as the Friedmann equation
and Eq.~(\ref{eq2}) as the acceleration equation in the literature,
whereas for simplicity we name both Eqs.~(\ref{eq1}) and (\ref{eq2})
Friedmann equations here. For generality, we assume that both
$\gamma$ and $\Lambda$ are functions of the scale factor $a$, thus
we call it the generalized FRW model. Combining the Friedmann
equations with the EOS, we obtain
\begin{equation}
\frac{\ddot{a}}{a}=-\frac{3\gamma(a)-2}{2}\frac{\dot{a}^2}{a^2}+\frac{\gamma(a)\Lambda(a)}{2},\label{eq:frw}
\end{equation}
which determines the evolution of the scale factor.

We regard Eq.~(\ref{eq:frw}) as a basic starting point; therefore,
if the dynamical equation for the scale factor can be written as
that form, the present framework can be valid. If the Newton
constant $G$ is constant and the cosmological constant $\Lambda$ is
variable, the energy-momentum tensor for the matter cannot
individually conserved \cite{vcc1,vcc2}, which implies an
interaction between the matter and vacuum energy. In the following,
we assume $G$ to be constant until Sec. IV.

Our aim is to find a Hamiltonian description of Eq.~(\ref{eq:frw})
as the classical equation of motion. We start from the following
Lagrangian
\begin{equation}
\mathcal{L}(q,\dot{q})=\frac{1}{2}M(q)\dot{q}^2-V(q),\label{eq:L}
\end{equation}
and the corresponding Hamiltonian thus is
\begin{equation}
\mathcal{H}(q,p)=\frac{p^2}{2M(q)}+V(q),\label{eq:H}
\end{equation}
with the canonical Poisson bracket $\{q,p\}=1$. One can check that
the equation of motion for Eq.~(\ref{eq:L}) or (\ref{eq:H}) is
\begin{equation}
\frac{\ddot{q}}{q}=-\frac{1}{2}\frac{\partial\ln M}{\partial\ln
q}\frac{\dot{q}^2}{q^2}-\frac{1}{Mq}\frac{\partial V}{\partial
q}.\label{eq:eom}
\end{equation}
This equation possesses the same form as Eq.~(\ref{eq:frw}).
Therefore, by comparing Eq.~(\ref{eq:frw}) with Eq.~(\ref{eq:eom}),
we can take $a$ as the general coordinate and solve the functions
$M(a)$ and $V(a)$. Then the Lagrangian
$\mathcal{L}=\frac{1}{2}M(a)\dot{q}^2-V(a)$ with
\begin{equation}
M=\exp\left(\int\frac{3\gamma-2}{a}da\right),\quad
V=-\frac{1}{2}\int M\gamma\Lambda ada,
\end{equation}
gives Eq.~(\ref{eq:frw}) as the equation of motion. For some
specified functions $\gamma=\gamma(a)$ and $\Lambda=\Lambda(a)$, the
above integrations can be evaluated out to give $M(a)$ and $V(a)$
explicitly.

Now we can see that the generalized FRW model essentially
corresponds to an object with variable mass $M(a)$ moving in a
potential field $V(a)$. In the following, we will show that this
picture can be further simplified as an object with constant mass
moving in an effective potential field $\tilde{V}(\phi)$, after an
appropriate transformation of the scale factor.

\subsection{Canonical transformation}
The above problem can be generalized as the Hamiltonian description
of the nonlinear equation
\begin{equation}
\ddot{q}=f_1(q)\dot{q}^2+f_2(q),\label{eq:ddq}
\end{equation}
where $f_1(q)$ and $f_2(q)$ are two specified functions. This
equation can be derived by the Lagrangian
$\mathcal{L}=\frac{1}{2}M(q)\dot{q}^2-V(q)$ with
\begin{equation}
M=\exp\left(-2\int f_1(q)dq\right),\quad V=-\int
Mf_2(q)dq.\label{eq:mv}
\end{equation}

We define a new variable $\phi$ as (see Appendix)
\begin{equation}
\phi\equiv\int\exp\left(-\int f_1(q)dq\right)dq.\label{eq:phi}
\end{equation}
This transformation can eliminate the $\dot{q}^2$ term and gives the
equation for the variable $\phi$ as in
\begin{equation}
\ddot{\phi}=\left.f_2(q)\exp\left(-\int
f_1(q)dq\right)\right|_{q\to\phi},\label{eq:ddp}
\end{equation}
where $q\to\phi$ denotes using Eq.~(\ref{eq:phi}) to change the
variable $q$ to $\phi$. Since there is no $\dot{\phi}^2$ term in
Eq.~(\ref{eq:ddp}), this can be regarded as a partial linearization.
Therefore, the system of Eq.~(\ref{eq:ddq}) transformed to
Eq.~(\ref{eq:ddp}) can be described by the Lagrangian
\begin{equation}
\mathcal{L}(\phi,\dot{\phi})=\frac{1}{2}\dot{\phi}^2-\tilde{V}(\phi),\label{eq:eff}
\end{equation}
with the potential as
\begin{eqnarray}
\tilde{V}(\phi) &=& -\int\left[f_2(q)\exp\left(-\int
f_1(q)dq\right)\right]_{q\to\phi}d\phi\nonumber\\
&=& \left.-\int f_2(q)\exp\left(-2\int
f_1(q)dq\right)dq\right|_{q\to\phi}.\label{eq:effV}
\end{eqnarray}
The simplification of the problem by Eq.~(\ref{eq:phi}) is
essentially the canonical transformation
\begin{equation}
q\to\phi,\quad p_q\to
p_\phi,\quad\mathcal{H}(q,p_q)\to\mathcal{H}(\phi,p_\phi),\label{eq:canon}
\end{equation}
where $p_q=M(q)\dot{q}$, $p_\phi=\dot{\phi}$, and
$\mathcal{H}(\phi,p_\phi)=\frac{1}{2}p_\phi^2+\tilde{V}(\phi)$.
Therefore, the classical and quantum properties of different models
are characterized by the effective potentials.

For Eq.~(\ref{eq:frw}) as a special case, the new variable $\phi$ is
given by
\begin{equation}
\phi=\int\exp\left(\int\frac{3\gamma-2}{2a}da\right)da.\label{eq:phia}
\end{equation}

\subsection{Some examples}
We will give some special cases of the above general framework to
show some applications. If both $\gamma$ and $\Lambda$ are constant
for a simple case, the integrations in Eq.~(\ref{eq:phia}) can be
evaluated out as
\begin{subequations}
\begin{eqnarray}
\phi &=& \frac{2}{3\gamma}a^{3\gamma/2},\quad\gamma\neq 0,\label{eq:pg}\\
&=& \ln a,\quad\gamma=0.
\end{eqnarray}
\end{subequations}
Now we consider $\gamma\neq 0$ for example. The special case
$\gamma=1$ corresponds to the $\Lambda$CDM model. The equation for
$\phi$ can be obtained as
$\ddot{\phi}-\frac{3}{4}\gamma^2\Lambda\phi=0$, and the
corresponding Lagrangian is
\begin{equation}
\mathcal{L}=\frac{1}{2}\dot{\phi}^2+\frac{3}{8}\gamma^2\Lambda\phi^2.\label{eq:lphi}
\end{equation}
We can see that the simplest model in cosmology just corresponds to
a harmonic oscillator after linearization. In particular, this is a
upside-down harmonic oscillator for the asymptotic de Sitter
Universe.

We can add the curvature effect to the $\Lambda$CDM model, which is
described by the special case $m=2$ of the following equation:
\begin{equation}
\frac{\ddot{a}}{a}=-\frac{3\gamma-2}{2}\frac{\dot{a}^2}{a^2}+\frac{\gamma\Lambda}{2}
-\frac{k}{a^m}.
\end{equation}
Here the parameters $\gamma$, $\Lambda$, and $m$ are all constants.
This equation possesses the same form of Eq.~(\ref{eq:ddq}). By
defining $\phi$ as Eq.~(\ref{eq:phi}) and using Eq.~(\ref{eq:effV}),
we obtain the effective potential as
\begin{equation}
\tilde{V}(\phi)=-\frac{3}{8}\gamma^2\Lambda\phi^2+\frac{k}{3\gamma-m}
\left(\frac{3\gamma\phi}{2}\right)^{2-2m/3\gamma},\label{eq:V}
\end{equation}
for $\gamma\neq 0$ and $m\neq 3\gamma$.

Another example is the Friedmann equations during the inflation era.
In the study of inflation, we usually use the conformal time $\tau$
instead of the comoving cosmic time $t$. Here we assume that a
constant term $-p_0$ is in the EOS during inflation. Thus the
Friedmann equations combined with the EOS $p=-\rho-p_0$ yield
\begin{equation}
\frac{a''}{a^3}=2\frac{a'^2}{a^4}+\frac{\kappa^2}{2}p_0,
\end{equation}
where the prime denotes a derivative with respect to $\tau$, and
$\kappa^2=8\pi G$. By defining $\phi=-1/a$, the equation for $\phi$
is $\phi''\phi+(\kappa^2/2)p_0=0$. The effective potential is thus
\begin{equation}
\tilde{V}(\phi)=\frac{\kappa^2}{2}p_0\ln|\phi|.
\end{equation}
Moreover, if we add the curvature term in this case, it corresponds
to a $\phi^2$ potential.

\section{Bulk viscosity}
We assume that the cosmic fluid possesses some dissipation effects.
Since the sheer tensor $\sigma_{\mu\nu}=0$ for RW metric, the sheer
viscosity does not contribute to the evolution in Friedmann
cosmology. The energy-momentum tensor for nonperfect fluid
concerning bulk viscosity in the right-hand side of Einstein's
equation is given by \cite{bre02,ren}
\begin{equation}
T_{\mu\nu}=\rho U_\mu U_\nu+(p-\zeta_0\theta)h_{\mu\nu},
\end{equation}
where $h_{\mu\nu}\equiv g_{\mu\nu}+U_\mu U_\nu$ is the projection
operator, $\theta\equiv U^{\alpha}_{;\alpha}=3\dot{a}/a$ is the
scalar expansion, and $\zeta$ is the bulk viscosity coefficient.
Consequently, Eq.~(\ref{eq:frw}) should be modified as
\begin{equation}
\frac{\ddot{a}}{a}=-\frac{3\gamma(a)-2}{2}\frac{\dot{a}^2}{a^2}+12\pi
G\zeta_0\frac{\dot{a}}{a}+\frac{\gamma(a)\Lambda(a)}{2}.\label{eq:vis}
\end{equation}
where both $\gamma$ and $\Lambda$ can be functions of $a$ for
generality, and $\zeta_0$ is constant. We also find a Hamiltonian
\begin{equation}
\mathcal{H}(a,p_a,t)=\frac{p_a^2}{2M(a,t)}+V(a,t),\label{eq:vish}
\end{equation}
with the Poisson bracket $\{a,p_a\}=1$ to give Eq.~(\ref{eq:vis}) as
the classical equation of motion. The functions in this Hamiltonian
are given by
\begin{subequations}
\begin{eqnarray}
M &=& \exp\left(\int\frac{3\gamma-2}{a}da-12\pi G\zeta_0 t\right),\\
V &=& -\frac{1}{2}\int M\gamma\Lambda ada.
\end{eqnarray}
\end{subequations}
Although a dissipative system cannot be described by a conservative
Hamiltonian generally, one can directly check that the classical
equation of motion for the Hamiltonian Eq.~(\ref{eq:vish}) is
Eq.~(\ref{eq:vis}). As a special case, the equation for a damping
harmonic oscillator can be derived by the Caldirora-Kani (CK)
Hamiltonian \cite{vis}.

The above problem can be generalized to construct a Hamiltonian
system for the equation
\begin{equation}
\ddot{q}=f_1(q)\dot{q}^2+\eta\dot{q}+f_2(q),\label{eq:gen}
\end{equation}
where $\eta$ is constant. It can be derived by the Hamiltonian
$H(q,p,t)=\frac{1}{2}M(q,t)^{-1}p^2+V(q,t)$ with
\begin{equation}
M=\exp\left(-2\int f_1(q)dq-\eta t\right),\quad V=-\int Mf_2(q)dq.
\end{equation}

Similarly, by using the new variable $\phi$ defined by
Eq.~(\ref{eq:phi}), the equation for $\phi$ is
\begin{equation}
\ddot{\phi}=\eta\dot{\phi}+\left.f_2(q)\exp\left(-\int
f_1(q)dq\right)\right|_{q\to\phi}.
\end{equation}
Now we consider a very special case that both $\gamma$ and $\Lambda$
are constant; then $\phi$ defined by Eq.~(\ref{eq:pg}) satisfies
\begin{equation}
\ddot{\phi}-12\pi
G\zeta_0\dot{\phi}-\frac{3}{4}\gamma^2\Lambda\phi=0,
\end{equation}
which describes a damping harmonic oscillator.

The damping harmonic oscillator
\begin{equation}
M\ddot{q}=-\eta\dot{q}(t)-\frac{\partial V(q)}{\partial
q},\label{eq:dho}
\end{equation}
has been studied in quantum mechanics. The CK Hamiltonian
\begin{equation}
H=\frac{1}{2M}e^{-\eta t/M}p^2+\frac{1}{2}M\omega^2 e^{\eta t/M}q^2,
\end{equation}
with the commutation relation $[q,p]=i\hbar$, can yield the
dissipation equation (\ref{eq:dho}) through the Heisenberg equation
\cite{vis}. Our work can be regarded as a generalization to the case
of variable mass. It is the variable mass that generates a nonlinear
term in the equation of motion that describes the generalized FRW
model.

In our previous work \cite{ren}, we have proposed an EOS as
\begin{equation}
p=(\gamma-1)\rho-\frac{2}{\sqrt{3}\kappa
T_1}\sqrt{\rho}-\frac{2}{\kappa^2 T_2^2}
\end{equation}
where the parameters $\gamma$, $T_1$ and $T_2$ are constants.
Combining the Friedmann equations with this more practical EOS, we
obtain the dynamical evolution equation for the scale factor as
\begin{equation}
\frac{\ddot{a}}{a}=-\frac{3\gamma-2}{2}\frac{\dot{a}^2}{a^2}+\frac{1}{T_1}\frac{\dot{a}}{a}+\frac{1}{T_2^2}.
\end{equation}
This model possesses a large variety of properties, such as that we
have found a scalar field model which is equivalent to the above
EOS. For related works on the modified EOS, see
Ref.~\cite{ren,odint,linear,sha00}. The present work can also be
regarded as a generalization of the EOS to $\gamma=\gamma(a)$ and
$T_2=T_2(a)$. And the corresponding Hamiltonian formalism for this
system can be constructed similarly.

\section{Relations to the observable quantities}
The observations of the supernovae (SNe) Ia have provided the direct
evidence for the cosmic accelerating expansion of our current
Universe \cite{de}. A bridge between the cosmological theory and the
observation data is the $H$-$z$ relation, where $H\equiv\dot{a}/a$
is the Hubble parameter and $z$ is the redshift. For example, the
$\Lambda$CDM model in cosmology can be described mainly as
$H^2(z)=H_0^2[\Omega_m(1+z)^3+1-\Omega_m]$, where $\Omega_m$ is the
matter energy density. This model fits the observational data well
and provides the cosmological constant as the simplest candidate for
dark energy. In a sense, the different cosmological models are
characterized by the corresponding $H$-$z$ relations.

There is also a systematic way to construct the Hamiltonian starting
from the general model
\begin{equation}
H^2=f(a),\label{eq:hf}
\end{equation}
where $f(a)$ is a specific function of the scale factor $a$,
according to the model. By differentiating Eq.~(\ref{eq:hf}), we
obtain that it is a solution of the following equation:
\begin{equation}
\frac{\ddot{a}}{a}=-\frac{3\gamma-2}{2}\frac{\dot{a}^2}{a^2}+\frac{3\gamma
f(a)}{2}+\frac{af'(a)}{2},
\end{equation}
which possesses the same form of Eq.~(\ref{eq:frw}) or
(\ref{eq:ddq}). The corresponding coefficients are given by
\begin{equation}
f_1(a)=-\frac{3\gamma-2}{2a},\quad f_2(a)=\frac{3\gamma
af(a)}{2}+\frac{a^2f'(a)}{2}.
\end{equation}
Then by applying Eq.~(\ref{eq:mv}) we can obtain the corresponding
Hamiltonian. Therefore, even if the EOS for a cosmological model is
not explicitly linear in $\rho$, the Hamiltonian formalism in the
present work can also be applied if the effective Friedmann equation
$H^2=f(a)$ can be given out for that model.

Many approaches such as modified gravity \cite{meng} can be reduced
to effective Friedmann equations in the form $H^2=f(a)$. Since
$\Lambda$CDM model fits the SNe Ia data well, the reasonable
cosmological models should be reduced to Friedmann cosmology in an
effective way and give out the right $H$-$z$ relation, in order to
make a comparison with the $\Lambda$CDM model. In our case, the
Friedmann equations in terms of the Hubble parameter can be written
as
\begin{equation}
aH\frac{dH}{da}=-\frac{3\gamma}{2}H^2+\tilde{\Lambda}(a).\label{eq:aH}
\end{equation}
Here $\gamma$ is assumed to be constant for simplicity. This
equation is linear in $H^2$ and the effective term
$\tilde{\Lambda}(a)$ is an inhomogeneous term. The solution in terms
of $H(z)$ concerning the initial condition $H(0)=H_0$ is given out
by
\begin{equation}
H(z)^2=H_0^2(1+z)^{3\gamma}\left[1-2\int_0^z\tilde{\Lambda}(z')(1+z')^{-3\gamma-1}dz'\right].
\end{equation}
In the power-law $\Lambda$CDM model, the contributions of different
components are separated in $H^2$, such as a constant for the
cosmological constant, and a $(1+z)^2$ factor for the curvature
term. But in the general case, the contribution of the matter cannot
be separated from the above solution. This problem is related to the
conservation law of the matter, which has been investigated in
Refs.~\cite{vcc1,vcc2}.

The Bianchi identity for the energy-momentum tensor
Eq.~(\ref{eq:tmn}) gives
\begin{equation}
\dot{\rho}_\Lambda+\dot{\rho}+3H(\rho+p)=0,
\end{equation}
which implies that energy transfer will exist between the matter and
the vacuum energy. An intuitive idea has been proposed that if both
$G$ and $\Lambda$ are variable, the ordinary energy-momentum tensor
can be individually conserved, i.e., $\dot{\rho}+3H(\rho+p)=0$
\cite{vcc2}. This is achieved by combining the Bianchi identity for
the variable $G$ and $\Lambda$ model
\begin{equation}
\frac{d}{dt}[G(\rho_\Lambda+\rho)]+3GH(\rho+p)=0,
\end{equation}
with the following constraint:
\begin{equation}
(\rho+\rho_\Lambda)\dot{G}+G\dot{\rho}_\Lambda=0.
\end{equation}
The authors of Ref.~\cite{vcc2} assume that both the Newton constant
$G$ and the cosmological constant $\Lambda$ are functions of a scale
parameter $\mu$ and apply the renormalization group approach to
cosmology. If $G(\mu)$ evolves by a logarithmic law and
$\rho_\Lambda(\mu)$ evolves quadratically with $\mu$, then this
picture can explain the evolution of the Universe, and at the same
time, the variable $G$ can explain the flat rotation curves of the
galaxies without introducing the dark matter hypothesis.

\section{Remarks on quantum cosmology}
We have obtained a classical Hamiltonian formalism of the Friedmann
equations. Generally, once a Hamiltonian is obtained, the system can
be quantized straightforwardly by replacing the Poisson bracket with
the commutation relation $[q,p]=i$. However, we need to take into
account the ambiguity in the ordering of noncommuting operators $q$
and $p$. For simplicity, we ignore the ordering ambiguity here. In
terms of the new variable $\phi$, the corresponding
Schr\"{o}dinger's equation can be written as
\begin{equation}
\mathcal{H}(\phi,\hat{p}_\phi)\Psi(\phi)=E\Psi(\phi),
\end{equation}
where $\hat{p}_\phi=-i\partial_\phi$. To make a comparison between
our approach and the Wheeler-DeWitt equation, we only take the
$\Lambda$CDM model as a very special case for an illustrative
example. The corresponding Hamiltonian for Eq.~(\ref{eq:lphi}) in
the case $\gamma=1$ is
\begin{equation}
\mathcal{H}=\frac{1}{2a}p^2-\frac{1}{6}\Lambda a^3,
\end{equation}
where $p=a\dot{a}$. In the approach of the Wheeler-DeWitt equation,
$\mathcal{H}=0$ is a constraint \cite{dewitt,vil94}, thus the
quantization gives $(\partial_a^2+\frac{\Lambda}{3}a^4)\Psi(a)=0$.
This is an anharmonic oscillator with zero energy eigenvalue. In our
case, the Hamiltonian is nonzero and proportional to the matter
energy density, which we show in the following. The solution of
Eq.~(\ref{eq:aH}) with $\tilde{\Lambda}(a)=\Lambda/2$ is
\begin{eqnarray}
H^2 &=& \left(H_0^2-\frac{\Lambda}{3}\right)a^{-3}+\frac{\Lambda}{3}\nonumber\\
&=& H_0^2[\Omega_ma^{-3}+1-\Omega_m],
\end{eqnarray}
where $\Omega_m\equiv 1-\Lambda/(3H_0^2)$. Therefore, the
Hamiltonian can be calculated as
\begin{equation}
\mathcal{H}=\frac{a^3}{2}\left(\frac{\dot{a}^2}{a^2}-\frac{\Lambda}{3}\right)=\frac{1}{2}H_0^2\Omega_m.
\end{equation}
After a canonical transformation by Eq.~(\ref{eq:canon}), the
Schr\"{o}dinger's equation in terms of $\phi$ becomes
\begin{equation}
\left[-\frac{1}{2}\frac{d^2}{d\phi^2}-\frac{3}{8}\Lambda\phi^2\right]\Psi(\phi)=E\Psi(\phi).
\end{equation}
Thus, for the asymptotic de Sitter Universe, the $\Lambda$CDM model
corresponds to an upside-down harmonic oscillator in our formalism.
Such an oscillator also appears in the matrix description of de
Sitter gravity \cite{gao01}.

We can transform the de Sitter Universe to the dual anti-de Sitter
Universe by employing the scale factor duality \cite{dual}, which
has been found that $a\to a^{-1}$ gives $H\to -H$ and other
consequences. The duality for Eq.~(\ref{eq:frw}) is given by
\begin{equation}
a\to a^{-1},\quad \gamma\to -\gamma,\quad \Lambda\to -\Lambda,\quad
\phi\to -\phi.
\end{equation}
It can be checked easily that Eq.~(\ref{eq:frw}) is invariant under
these transformations. If we use the dual scale factor $a^{-1}$, the
corresponding potential becomes
$\tilde{V}(\phi)=+\frac{3}{8}\gamma^2\Lambda\phi^2$. In fact,
quantization in de Sitter spacetime is one of the major difficulties
of string theory at one time (though this picture has changed a
little bit after Kachru-Kallosh-Linde- Trivedi theory appeared). It
seems that quantizing de Sitter cosmology is no difference, since
the time variable used is the same, and it is known that there is no
global timelike coordinates in de Sitter spacetime. Some quantum
effects of a scalar field in de Sitter background can be found in
Ref.~\cite{vko}.

\section{Conclusion and discussion}
We have proposed a systematic scheme to describe the Friedmann
equations through a Hamiltonian formalism. The generalized FRW model
accompanied by both variable EOS parameter and variable cosmological
constant admits a Hamiltonian description without constraint. After
an appropriate canonical transformation, the system can be
significantly simplified to an object moving in an effective
potential field. The bulk viscosity can also be taken into account
by a time-dependent Hamiltonian. Some examples are given explicitly,
such as the $\Lambda$CDM model, the curvature term effect, and the
inflation period. The quantization of the system provides a new
approach to study the potential quantum cosmology, which is an
intriguing topic in theoretical physics research.

We shall discuss some possible future developments of our work. As
we have claimed, the formalism in this work can be applied to a
large variety of cosmological models. By solving the Schr\"{o}dinger
equation $\mathcal{H}(\phi,\hat{p}_\phi)\Psi=E\Psi$, the
cosmological wave function can be obtained for a specific model.
Here we consider the curvature effect, for example, which is
described by the potential Eq.~(\ref{eq:V}) with parameters
$\Lambda=0$, $\gamma=1$, and $m=2$. The corresponding
Schr\"{o}dinger equation can be solved in terms of the biconfluent
Heun equation (BHE) \cite{slav}. We can also start from the
effective Lagrangian and study the observational effects when we
modify the potential. We believe that our formalism would give a new
perspective to the potential study of quantum cosmology physics.

\section*{ACKNOWLEDGMENTS}
J.R. thanks Prof. M.L. Ge for helpful discussions on Hamiltonian
systems. X.H.M. is supported by NSFC under No. 10675062 and BK21
Foundation. L.Z. is supported by NSFC under No. 90403014.

\appendix
\section{Mathematical notes}
A more general correspondence between a Hamiltonian and its equation
of motion is given in Ref.~\cite{slav}. The equation of motion of
the Hamiltonian
\begin{equation}
H(q,p,t)=\frac{1}{f(t)}\left(P_0(q,t)p^2+P_1(q,t)p+P_2(q,t)\right),
\end{equation}
is given out by
\begin{eqnarray}
\ddot{q} &=& \frac{1}{2}\frac{\partial\ln P_0}{\partial
q}\frac{\dot{q}^2}{q^2}-\left(\frac{\partial\ln f}{\partial
t}-\frac{\partial\ln P_0}{\partial
t}\right)\dot{q}\nonumber\\
&& +\frac{P_0}{f^2}\left(\frac{\partial}{\partial
q}\frac{P_1^2}{2P_0}+f\frac{\partial}{\partial
t}\frac{P_1}{P_0}-2\frac{\partial V}{\partial
q}\right).\label{eq:slav}
\end{eqnarray}
In the mathematical aspect, Eq.~(\ref{eq:gen}) can be further
generalized to the following equation:
\begin{equation}
\ddot{q}=F_1(q,t)\dot{q}^2+F_2(q,t)\dot{q}+F_3(q,t),
\end{equation}
however, here the coefficients $F_1$ and $F_2$ are not completely
independent. Comparing with Eq.~(\ref{eq:slav}), we can see that the
condition $2\partial_t F_1(q,t)=\partial_q F_2(q,t)$ must be
satisfied for consistency. In the present work, both $f_1(q)$ and
$\eta$ have safely satisfied this condition.

We shall explain why we choose the transformation as in
Eq.~(\ref{eq:phi}). Starting from the following equation
\begin{equation}
\ddot{q}=f_1(q)\dot{q}^2+\eta\dot{q}+f_2(q),
\end{equation}
we expect that after an appropriate change of variable $\phi(q)$,
the above equation can be transformed as
\begin{equation}
\ddot{\phi}=\eta\dot{\phi}+g(\phi).\label{eq:a}
\end{equation}
By differentiating $\phi(q)$, we obtain $\dot{\phi}=\phi'\dot{q}$,
and $\ddot{\phi}=\phi''\dot{q}^2+\phi'\ddot{q}$, where the prime
denotes a derivative with respect to $q$. Substituting $\phi$,
$\dot{\phi}$, and $\ddot{\phi}$ into Eq.~(\ref{eq:a}), we obtain
\begin{equation}
\ddot{q}=-\frac{\phi''}{\phi'}\dot{q}^2+\eta\dot{q}+\frac{g}{\phi'}
\end{equation}
Now it turns out that by defining $-\phi''/\phi'=f_1(q)$, which can
be solved as the form Eq.~(\ref{eq:phi}), the $\dot{q}^2$ term can
be eliminated.

\end{document}